\documentclass[a4paper,11pt]{article}
\usepackage[dvips]{graphicx}
\usepackage{amssymb}
\usepackage{amsmath}

\begin{document}

\title{Witten Index and Superconducting Strings}
\author{V.K.Oikonomou\thanks{
voiko@physics.auth.gr}\\
Technological Education Institute of Serres, \\
Dept. of Informatics and Communications 62124 Serres, Greece\\
and\\
Dept. of Theoretical Physics Aristotle University of Thessaloniki,\\
Thessaloniki 541 24 Greece} \maketitle

\begin{abstract}
The Yukawa interaction sector of superstring inspired models that
give superconducting strings, can be described in terms of a
supersymmetric quantum mechanics algebra. We relate the Witten
index of susy quantum mechanics with an index characteristic to
superconducting string models.
\end{abstract}

\section*{Introduction}

Superconducting strings are known to have important cosmological
implications \cite{witten,supercondstrings}. Cosmic strings can
become superconducting if charged fermionic transverse zero modes
are trapped along the strings \cite{rossi}. For example in
\cite{rossi} a single massive fermion was considered which
acquired its mass through a Yukawa-type interaction with a scalar
field having varying phase around the string. In \cite{weinberg}
an index theorem was obtained, which determines the minimum number
of zero modes.

\noindent Moreover in \cite{ganoulis} an index theorem was
developed, which applies in more realistic theories. Particularly
in grand-unified or superstring inspired models one has many
left-right handed fermions coupled to a number of scalar fields
with Yukawa interactions. Some of these models admit cosmic
strings solutions and it is interesting to know which of these are
superconducting. The index theorem developed in reference
\cite{ganoulis} gives an adequate solution to this and applies to
models where one has a matrix of Higgs fields and many charged
fermion flavors coupled to this matrix with arbitrary phase
variations around the string. Moreover the nonzero index case (we
denote the index $I_q$) is a criterion whether the cosmic strings
are superconducting or not.

\noindent In this letter we shall relate the index $I_q$ with the
Witten index of supersymmetric quantum mechanical systems. Indeed
we shall see that models that admit superconducting string
solutions can be written in terms of a $N=2$ supersymmetric
quantum mechanics systems and also that the Witten index of this
system is identical to the $I_q$ index. Thus we relate a purely
mathematical property of a system to the phenomenology of a
grand-unified or superstring inspired model.

\noindent We shall briefly present some features of supersymmetric
quantum mechanics and also the required background for
superconducting strings and the index $I_q$, in order to make the
article self contained.

\section*{Supersymmetric Quantum Mechanics and Superconducting Strings}

\subsection*{Supersymmetric Quantum Mechanics}

Let us briefly review here some properties of to supersymmetric
quantum mechanics. The presentation is based on \cite{susyqm}. A
quantum system, described by a Hamiltonian $H$, which is
characterized by the set $\{H,Q_1,...,Q_N\}$, with $Q_i$ self
adjoint operators, is called supersymmetric if the following
anti-commutation relation holds for $i=1,2,...N$,
\begin{equation}\label{susy1}
\{Q_i,Q_j\}=H\delta_{i{\,}j}
\end{equation}
The self-adjoint operators are then called supercharges and the
Hamiltonian ``$H$" is called SUSY Hamiltonian. The algebra
(\ref{susy1}) describes a symmetry called N-extended
supersymmetry. Of course SUSY quantum mechanics can be defined in
terms of non self-adjoint supercharges, as we will see shortly.
The superalgebra (\ref{susy1}) poses some restrictions on the SUSY
Hamiltonian, particularly it follows due to the anti-commutation
that,
\begin{equation}\label{susy3}
H=2Q_1^2=Q_2^2=\ldots =2Q_N^2=\frac{2}{N}\sum_{i=1}^{N}Q_i^2.
\end{equation}
A supersymmetric quantum system $\{H,Q_1,...,Q_N\}$ is said to
have good susy (unbroken supersymmetry) if its ground state
vanishes, that is $E_0=0$. For a positive ground-state energy with
$E_0>0$, susy is said to be broken. It is obvious that for good
supersymmetry, the Hilbert space eigenstates must be annihilated
by all supercharges, that is,
\begin{equation}\label{s1}
Q_i |\psi_0^j\rangle=0
\end{equation}
for all $i,j$. We now describe the basic features of $N=2$
supersymmetric quantum mechanics. The $N=2$ algebra consists of
two supercharges $Q_1$ and $Q_2$ and a Hamiltonian $H$, which obey
the following relations,

\begin{equation}\label{sxer2}
\{Q_1,Q_2\}=0,{\,}{\,}{\,}H=2Q_1^2=2Q_2^2=Q_1^2+Q_2^2
\end{equation}
A more frequently used notation involves the following operators,
\begin{equation}\label{s2}
Q=\frac{1}{\sqrt{2}}(Q_{1}+iQ_{2})
\end{equation}
and the adjoint,
\begin{equation}\label{s255}
Q^{\dag}=\frac{1}{\sqrt{2}}(Q_{1}-iQ_{2})
\end{equation}
The operators of relations (\ref{s2}) and (\ref{s255}) satisfy the following equations,
\begin{equation}\label{s23}
Q^{2}={Q^{\dag}}^2=0
\end{equation}
and also can be written in terms of the Hamiltonian as,
\begin{equation}\label{s4}
\{Q,Q^{\dag}\}=H
\end{equation}
It is always possible for $N=2$ to define the Witten parity operator, $W$, which is defined through the following relations,
\begin{equation}\label{s45}
[W,H]=0
\end{equation}
and
\begin{equation}\label{s5}
\{W,Q\}=\{W,Q^{\dag}\}=0
\end{equation}
Also $W$ satisfies,
\begin{equation}\label{s6}
W^{2}=0
\end{equation}
Using $W$, we can span the Hilbert space $\mathcal{H}$ of the
quantum system to positive and negative Witten-parity spaces,
defined as, $\mathcal{H}^{\pm}=P^{\pm}\mathcal{H}=\{|\psi\rangle :
W|\psi\rangle=\pm |\psi\rangle $. Thus the Hilbert space
$\mathcal{H}$ is decomposed into the eigenspaces of $W$, so
$\mathcal{H}=\mathcal{H}^+\oplus \mathcal{H}^-$, and each operator
acting on the vectors of $\mathcal{H}$ is represented in general
by $2N\times 2N$ matrices. We shall use the representation
\begin{equation}\label{s7345}
W=\bigg{(}\begin{array}{ccc}
  I & 0 \\
  0 & -I  \\
\end{array}\bigg{)}
\end{equation}
with $I$ the $N\times N$ identity matrix. Bearing in mind that $Q^2=0$ and $\{Q,W\}=0$, the supercharges are necessarily of the form,
\begin{equation}\label{s7}
Q=\bigg{(}\begin{array}{ccc}
  0 & A \\
  0 & 0  \\
\end{array}\bigg{)}
\end{equation}
and
\begin{equation}\label{s8}
Q^{\dag}=\bigg{(}\begin{array}{ccc}
  0 & 0 \\
  A^{\dag} & 0  \\
\end{array}\bigg{)}
\end{equation}
which imply,
\begin{equation}\label{s89}
Q_1=\frac{1}{\sqrt{2}}\bigg{(}\begin{array}{ccc}
  0 & A \\
  A^{\dag} & 0  \\
\end{array}\bigg{)}
\end{equation}
and also,
\begin{equation}\label{s10}
Q_2=\frac{i}{\sqrt{2}}\bigg{(}\begin{array}{ccc}
  0 & -A \\
  A^{\dag} & 0  \\
\end{array}\bigg{)}
\end{equation}
The $N\times N$ matrices $A$ and $A^{\dag}$ are generalized
annihilation and creation operators. Particularly $A$ acts as
follows, $A: \mathcal{H}^-\rightarrow \mathcal{H}^+$ and
$A^{\dag}$ as, $A^{\dag}: \mathcal{H}^+\rightarrow \mathcal{H}^-$
In the representation (\ref{s7345}), (\ref{s7}), (\ref{s8}) the
quantum mechanical Hamiltonian $H$, can be written in the diagonal
form,
\begin{equation}\label{s11}
H=\bigg{(}\begin{array}{ccc}
  AA^{\dag} & 0 \\
  0 & A^{\dag}A  \\
\end{array}\bigg{)}
\end{equation}
Thus for a $N=2$ supersymmetric quantum system, the total supersymmetric Hamiltonian $H$, consists of two superpartner Hamiltonians,
\begin{equation}\label{h1}
H_{+}=A{\,}A^{\dag},{\,}{\,}{\,}{\,}{\,}{\,}{\,}H_{-}=A^{\dag}{\,}A
\end{equation}
The above two Hamiltonians are known to be isospectral for
eigenvalues different from zero, that is,
\begin{equation}\label{isosp}
\mathrm{spec}(H_{+})\setminus \{0\}=\mathrm{spec}(H_{-})\setminus
\{0\}
\end{equation}
The eigenstates of $P^{\pm}$ are called positive and negative
parity eigenstates and are denoted as $|\psi^{\pm}\rangle$, with,
\begin{equation}\label{fd1}
P^{\pm}|\psi^{\pm}\rangle =\pm |\psi^{\pm}\rangle
\end{equation}
In the representation (\ref{s7345}), the parity eigenstates are
represented in the form,
\begin{equation}\label{phi5}
|\psi^{+}\rangle =\left(%
\begin{array}{c}
  |\phi^{+}\rangle \\
  0 \\
\end{array}%
\right)
\end{equation}
and also,
\begin{equation}\label{phi6}
|\psi^{-}\rangle =\left(%
\begin{array}{c}
  0 \\
  |\phi^{-}\rangle \\
\end{array}%
\right)
\end{equation}
with $|\phi^{\pm}\rangle$ $\epsilon$ $H^{\pm}$.

\noindent Let us now see which are the ground state properties for
good supersymmetry. For good supersymmetry as we noted before,
there exists at least one state in the Hilbert space with
vanishing energy eigenvalue, that is $H|\psi_{0}\rangle =0$. Since
the Hamiltonian commutes with the supercharges, $Q$ and
$Q^{\dag}$, it is obvious that, $Q|\psi_{0}\rangle =0$ and
$Q^{\dag}|\psi_{0}\rangle =0$. For a negative parity ground state,
\begin{equation}\label{phi5}
|\psi^{-}_0\rangle =\left(%
\begin{array}{c}
  |\phi^{-}_{0}\rangle \\
  0 \\
\end{array}%
\right)
\end{equation}
this implies that $A|\phi^{-}_{0}\rangle =0$, whereas for a
negative parity ground state,
\begin{equation}\label{phi6s6}
|\psi^{+}_{0}\rangle =\left(%
\begin{array}{c}
  0 \\
  |\phi^{+}_0\rangle \\
\end{array}%
\right)
\end{equation}
it implied that $A^{\dag}|\phi^{+}_{0}\rangle =0$. In general a
ground state can have positive or negative Witten parity and when
the ground state is degenerate both cases can occur. When $E\neq
0$ the number of positive parity eigenstates is equal to the
negative parity eigenstates. This does not happen for the ground
states. A rule to decide if there are zero modes is the so called
Witten index. Let $n_{\pm}$ the number of zero modes of $H_{\pm}$
in the subspace $\mathcal{H}^{\pm}$. For finite $n_{+}$ and
$n_{-}$ the quantity,
\begin{equation}\label{phil}
\Delta =n_{-}-n_{+}
\end{equation}
is called the Witten index. Whenever the Witten index is non-zero
integer, supersymmetry is good (unbroken). If the Witten index is
zero, it is not clear whether supersymmetry is broken (which would
mean $n_{+}=n_{-}=0$) or not ($n_{+}\neq n_{-}\neq 0$). The Witten
index is related to the Fredholm index of the operator $A$ we
mentioned earlier as,
\begin{equation}\label{ker}
\mathrm{ind} A = \mathrm{dim}{\,}\mathrm{ker}
A-\mathrm{dim}{\,}\mathrm{ker} A^{\dag}=
\mathrm{dim}{\,}\mathrm{ker}A^{\dag}A-\mathrm{dim}{\,}\mathrm{ker}AA^{\dag}
\end{equation}
The importance of the Fredholm index is that it is a topological
invariant. We shall use only Fredholm operators. For a discussion
on non-Fredholm operators and the Witten index, see \cite{susyqm}.
The Witten index is obviously related to the Fredholm index of
$A$, as,
\begin{equation}\label{ker1}
\Delta=\mathrm{ind} A=\mathrm{dim}{\,}\mathrm{ker}
H_{-}-\mathrm{dim}{\,}\mathrm{ker} H_{+}
\end{equation}

\subsection*{Superconducting Strings}

We now briefly present the theory of superconducting strings in
terms of Yukawa interactions of left-handed and right-handed
fermions with Higgs scalars. We follow closely reference
\cite{ganoulis}. Consider a theory containing $N$ left handed
fermion fields $\psi_{\alpha}$ and $N$ right-handed fermions
$\chi_{\alpha}$, interacting with the Higgs sector according to
the following Lagrangian,
\begin{equation}\label{lagrangian}
\mathcal{L}=i
\bar{\psi_{\alpha}}\gamma^{\mu}\partial_{\mu}\psi_{\alpha}+i
\bar{\chi_{\alpha}}\gamma^{\mu}\partial_{\mu}\chi_{\alpha}-(\bar{\chi_{\alpha}}M_{\alpha
\beta}\psi_{\beta}+\mathrm{H.c}).
\end{equation}
with $\alpha ,\beta =1,...,N$. The $N\times N$ matrix $M$ contains
the scalar fields with the interaction couplings. In general in a
string background the matrix $M$ depends only on the polar
coordinates $r$ and $\theta$ around the string. Due to the
cylindrical symmetry of the string, the theory has effectively two
dimensions and we can work in terms of two component spinors. The
chiral fermions can be written,
\begin{equation}\label{psibars}
\psi_{\alpha}=\frac{1}{\sqrt{2}}\left(%
\begin{array}{c}
  \widehat{\psi_{\alpha}} \\
  -\widehat{\psi_{\alpha}} \\
\end{array}%
\right)
\end{equation}
and also,
\begin{equation}\label{psibars1}
\chi_{\alpha}=\frac{1}{\sqrt{2}}\left(%
\begin{array}{c}
  \widehat{\chi_{\alpha}} \\
  -\widehat{\chi_{\alpha}} \\
\end{array}%
\right)
\end{equation}
Using an appropriate representation for the $\gamma$-matrices, the
Lagrangian can be written,
\begin{align}\label{yeaaah}
&\mathcal{L}=i\widehat{\psi_{\alpha}}^{\dag}\partial_{0}\psi_{\alpha}-i\widehat{\psi_{\alpha}}^{\dag}\sigma^{j}\partial_{j}\psi_{\alpha}
i\widehat{\chi_{\alpha}}^{\dag}\partial_{0}\chi_{\alpha}-i\widehat{\chi_{\alpha}}^{\dag}\sigma^{j}\partial_{j}\chi_{\alpha}\notag
\\&  - \widehat{\chi_{\alpha}^{\dag}}M_{\alpha
\beta}\widehat{\psi_{\alpha}}-\widehat{\psi_{\alpha}}^{\dag}M_{\alpha
\beta}\widehat{\chi_{\alpha}}
\end{align}
The equations of motion corresponding to the Lagrangian
(\ref{yeaaah}) are,
\begin{align}\label{ref1}
&-\partial_{0}\widehat{\psi_{\alpha}}+\sigma^{j}\partial_{j}\widehat{\psi_{\alpha}}-iM_{\alpha
\beta}^{\dag}\widehat{\chi_{\beta}}=0\\&\notag
-\partial_{0}\widehat{\chi_{\alpha}}+\sigma^{j}\partial_{j}\widehat{\chi_{\alpha}}-iM_{\alpha
\beta}^{\dag}\widehat{\psi_{\beta}}=0
\end{align}
with $\alpha,\beta=1,2,...,n$, and $\sigma^j$ the Pauli matrices.
Set,
\begin{equation}\label{ref2}
\widehat{\psi_{\alpha}}=f(x_3,t)\left(%
\begin{array}{c}
  \psi_{\alpha}(r,\phi) \\
  0 \\
\end{array}%
\right)
\end{equation}
and also,
\begin{equation}\label{ref2}
\widehat{\chi_{\alpha}}=f(x_3,t)\left(%
\begin{array}{c}
  0 \\
  \chi_{\alpha}(r,\phi) \\
\end{array}%
\right)
\end{equation}
Using the above two, the transverse zero-mode equations in the
$x_1{\,}x_2$ plane, read,
\begin{align}\label{koryfaia}
&(\partial_{1}+i\partial_{2})\psi_{\alpha}-iM_{\alpha
\beta}^{\dag}\chi_{\beta}=0\\&\notag
(\partial_{1}-i\partial_{2})\chi_{\alpha}+iM_{\alpha
\beta}\psi_{\beta}=0
\end{align}
Additionally one must have,
\begin{equation}\label{yeah}
(\partial_{0}-\partial_{3})f=0
\end{equation}
The last equation means that both $\psi$ and $\chi$ are left
movers (L-movers, see \cite{witten}). Another possibility is to
have,
\begin{equation}\label{ref2}
\widehat{\psi_{\alpha}}=f(x_3,t)\left(%
\begin{array}{c}
  0 \\
  \psi_{\alpha}(r,\phi) \\
\end{array}%
\right)
\end{equation}

\begin{equation}\label{ref2wer}
\widehat{\chi_{\alpha}}=f(x_3,t)\left(%
\begin{array}{c}
  \chi_{\alpha}(r,\phi) \\
  0 \\
\end{array}%
\right)
\end{equation}
with corresponding equations of motion,
\begin{align}\label{koryfaiax}
&(\partial_{1}-i\partial_{2})\psi_{\alpha}-iM_{\alpha
\beta}^{\dag}\chi_{\beta}=0\\&\notag
(\partial_{1}+i\partial_{2})\chi_{\alpha}+iM_{\alpha
\beta}\psi_{\beta}=0
\end{align}
and also,
\begin{equation}\label{yeah345}
(\partial_{0}+\partial_{3})f=0
\end{equation}
In this case both $\psi$ and $\chi$ are right movers (R-movers,
see \cite{witten}). The main interest in these theories is focused
on the above zero modes. For a general mass matrix $M_{\alpha
\beta}$, the solutions of (\ref{koryfaia}) and (\ref{koryfaiax})
are difficult to find. We define,
\begin{equation}\label{dmatrix}
\mathcal{D}=\left(%
\begin{array}{cc}
  \partial_{1}+i\partial_{2} & -iM^{\dag} \\
  iM & \partial_{1}-i\partial_{2} \\
\end{array}%
\right)_{2N\times 2N}
\end{equation}
and additionally,
\begin{equation}\label{dmatrix}
\mathcal{D}^{\dag}=\left(%
\begin{array}{cc}
  \partial_{1}-i\partial_{2} & -iM^{\dag} \\
  iM & \partial_{1}+i\partial_{2} \\
\end{array}%
\right)_{2N\times 2N}
\end{equation}
acting on
\begin{equation}\label{wee}
\left(%
\begin{array}{c}
  \psi_{\alpha} \\
  \chi_{\alpha} \\
\end{array}%
\right)
\end{equation}
The solutions of (\ref{koryfaia}) and (\ref{koryfaiax}) are the
zero modes of $D$ and also $D^{\dag}$. The Fredholm index $I_q$ of
the operator $\mathcal{D}^{\dag}$, is equal to,
\begin{equation}\label{indexd}
\mathrm{ind}D=\mathrm{I}_q=\mathrm{dim{\,}ker}(D^{\dag})-\mathrm{dim{\,}ker}(D)
\end{equation}
which is the number of zero modes of $\mathcal{D}$ minus the
number of zero modes of $\mathcal{D}^{\dag}$ and equals to the
number of the right movers $R$ minus the number of the left movers
$L$. The mass matrix is assumed to have the following form,
\begin{equation}\label{indexofd}
M_{\alpha \beta}(r,\phi)=S_{\alpha \beta}(r)e^{iq_{\alpha
\beta}\phi}
\end{equation}
The integers $q_{\alpha \beta}$ are related to the charges of the
fields with respect to the group generator $Q$ which corresponds
to the string \cite{witten}. With $\bar{q}_{\alpha}$ and
$q_{\beta}$ the charges of the fermion fields
$\chi_{\alpha}^{\dag}$ and $\psi_{\beta}$, the neutrality of
$\chi_{\alpha}^{\dag}M_{\alpha \beta}\psi_{\beta}$ implies
\begin{equation}\label{tetanus}
q_{\alpha \beta}=\bar{q}_{\alpha}-q_{\beta}
\end{equation}

\noindent It is proved that $I_q=\sum_{\alpha =1}^{n}q_{\alpha
\alpha}$ \cite{weinberg,ganoulis}. Therefore the Fredholm index of
$D$ is related to the charges of the fermions to the string gauge
group.

\noindent We can see that the theory of superconducting string
zero modes, defines a $N=2$ supersymmetric quantum mechanical
system. Indeed we can write,
\begin{equation}\label{wit2}
Q=\bigg{(}\begin{array}{ccc}
  0 & D \\
  0 & 0  \\
\end{array}\bigg{)}
\end{equation}
and additionally,
\begin{equation}\label{wit3}
Q^{\dag}=\bigg{(}\begin{array}{ccc}
  0 & 0 \\
  D^{\dag} & 0  \\
\end{array}\bigg{)}
\end{equation}
Also the Hamiltonian of the system can be written,
\begin{equation}\label{wit4}
H=\bigg{(}\begin{array}{ccc}
  DD^{\dag} & 0 \\
  0 & D^{\dag}D  \\
\end{array}\bigg{)}
\end{equation}
It is obvious that the above matrices obey, $\{Q,Q^{\dag}\}=H$,
$Q^2=0$, ${Q^{\dag}}^2=0$, $\{Q,W\}=0$, $W^2=I$ and $[W,H]=0$.
Thus we can relate the Witten index of the $N=2$ supersymmetric
quantum mechanics system with the index $I_q$ of the charges that
the fermions have. Indeed we have $I_q=-\Delta$, because,
\begin{equation}\label{ker}
I_q=\mathrm{dim}{\,}\mathrm{ker}
D^{\dag}-\mathrm{dim}{\,}\mathrm{ker} D=
\mathrm{dim}{\,}\mathrm{ker}DD^{\dag}-\mathrm{dim}{\,}\mathrm{ker}D^{\dag}D=-\mathrm{ind}D=-\Delta=n_--n_+
\end{equation}
So it is clear that the underlying supersymmetric algebra is
related to the phenomenology of the model on which the
superconducting string is based. This is very valuable because one
can answer the question if a model gives superconducting string
solution by examining the Witten index of the corresponding $N=2$
supersymmetric algebra. Before proceeding to some examples, let us
discuss some important issues. Due to the supersymmetric quantum
mechanical structure of the system, the zero modes of the operator
$D$ are related to the zero modes of the operator $DD^{\dag}$.
Therefore we can say that the zero modes of $DD^{\dag}$ and
$D^{\dag}D$ can be classified according to the Witten parity, to
parity positive and parity negative solutions. The last is
valuable in order to find solution to the equations
(\ref{koryfaia}) and (\ref{koryfaiax}). It is known
\cite{ganoulis} that when $I_q\neq 0$ then string
superconductivity is guaranteed. According to relation
(\ref{ker}), string superconductivity occurs when the Witten index
$\Delta$ is non-zero (susy unbroken). So when supersymmetry is not
broken, the theory we described admits superconducting solutions.
Also when the theory admits superconducting solutions (R-movers
and L-movers) supersymmetry is good-unbroken. However according to
\cite{ganoulis} when $I_q$ it is not sure whether superconducting
strings exist or not. Actually there may be some cases in which
solutions exist, while $I_q=0$. Does this means that the number of
R-movers is equal to L-movers or there are no zero modes? It is
found in \cite{ganoulis} that when someone uses the index $I_q$ it
is a good criterion to deal with these problems. Therefore we can
decide if supersymmetry is broken or not.

\noindent Let us give an example at this point (we follow
\cite{ganoulis}). Consider a superstring inspired model based on a
subgroup $G$ of $E_6$, which has an additional $U(1)$ factor along
with the Standard Model group, that is $G=SU(3)_c\times
SU(2)_L\times U(1)_Y\times U(1)_{L-R}$. The breaking of
$U(1)_{L-R}$ gives rise to cosmic strings. These models contain
singlets under $SU(3)_c\times SU(2)_L\times U(1)_Y$, which are
responsible for the breaking of the additional $U(1)_{L-R}$. When
one performs a non-trivial $L-R$ transformation, the
$SU(3)_c\times SU(2)_L\times U(1)_Y$ singlets acquire a phase
around the string. These fields are, $S_i=\langle S_i\rangle
e^{i\phi}$, $\tilde{S_i}=\langle S_i\rangle e^{i\phi}$
($i=1,2,3$), $N=\langle N\rangle e^{i\phi}$. The field
$\tilde{S_i}$ is the mirror of $S_i$. The field $N$ does not have
a mirror. The model contains the Higgs doublets, $H=\langle
H\rangle e^{in\phi}$ and also $\tilde{H}=\langle \tilde{H}\rangle
e^{i(n+1)\phi}$, $n=$integer. If we examine the down quark mass
matrix, ignoring fermion states from incomplete multiplets, the
mass matrix is,
\begin{equation}\label{naai}
\begin{array}{ccc}
    &\begin{array}{ccccc}
     g & & & & Q_D \\
      & & & & \\
    \end{array}\\
  M=\begin{array}{c}
    g_c \\
    D_c \\
  \end{array} & \left(%
\begin{array}{c|c}
 \langle S_1\rangle & 0 \\
  \hline
\langle N\rangle e^{i\phi} & \langle H\rangle e^{in\phi} \\
\end{array}%
\right)_{18\times 18} \\
\end{array}
\end{equation}
The interactions that give rise to the above mass matrix are,
$gg_cS$, $gD_cN$ and $Q_DD_cN$. The heavy quark states $g$, and
$g_c$ mix with the $d$-quark states $D_c$ and $Q_D$. The fermion
families are 3 and each flavor has 3 colors so each block has
$9\times 9$ dimension. In the above when $n\neq -1$, then $I_q\neq
0$ (and actually $I_q=1+n$ according to $I_q=\sum_{\alpha
=1}^{n}q_{\alpha \alpha}$) the cosmic strings are superconducting.
Thus in this case the scalar-fermion sector has Witten index
$\Delta \neq 0$. Therefore the quantum mechanical supersymmetry is
unbroken (good supersymmetry). However when $n=-1$, then $I_q=0$,
nevertheless according to \cite{ganoulis}, there are $9$ L-movers
and $9$ R-movers. So someone could say that supersymmetry is good
(unbroken), and so the positive parity states are equal to
negative parity states.

\end{document}